# NMR profiling of quantum electron solids in high magnetic fields

L. Tiemann[1,2,*], T. D. Rhone[1,2,*], N. Shibata[3] and K. Muraki[1,2]

[1]NTT Basic Research Laboratories, NTT Corporation, 3-1 Morinosato-Wakamiya, Atsugi 243-0198, Japan.

[2]ERATO Nuclear Spin Electronics Project, Japan Science and Technology Agency (JST), Kawaguchi 332-0012, Japan.

[3]Department of Physics, Tohoku University, Sendai 980-8578, Japan

---

[*] These authors contributed equally to this work.

When the motion of electrons is restricted to a plane under a perpendicular magnetic field *B*, a variety of quantum phases emerge at low temperatures whose properties are dictated by the Coulomb interaction and its interplay with disorder. At very strong *B*, the sequence of fractional quantum Hall (FQH) liquid phases[1] terminates in an insulating phase, which is widely believed to be due to the solidification of electrons into domains possessing Wigner crystal[2] (WC) order[3-11]. The existence of such WC domains is signaled by the emergence of microwave pinning-mode resonances[10,11], which reflect the mechanical properties characteristic of a solid. However, the most direct manifestation of the broken translational symmetry accompanying the solidification—the spatial modulation of particles' probability amplitude—has not been observed yet. Here, we demonstrate that nuclear magnetic resonance (NMR) provides a direct probe of the density topography of electron solids in the integer and fractional quantum Hall regimes. The data uncover quantum and thermal fluctuation of lattice electrons resolved on the nanometre scale. Our results pave the way to studies of other exotic phases with non-trivial spatial spin/charge order.

Wigner crystallisation is a concept originally proposed to occur in a disorder-free dilute electron system at zero magnetic field, when the Coulomb energy dominates over the kinetic energy[2]. Application of a strong perpendicular $B$ to a two-dimensional electron system (2DES) also facilitates Wigner crystallisation[12-15], as it quantises the electron energy into a discrete spectrum known as Landau levels (LLs) and thereby quenches the kinetic energy. As the electrons are confined to orbits with spatial extent of the order of the magnetic length $\ell_B = (h/2\pi eB)^{1/2}$ ($e$: elementary charge, $h$: Planck's constant), Wigner crystallisation becomes governed not by the electron density $n$, but by the LL filling factor $\nu = nh/eB$ ($= 2\pi n\ell_B^2$). Evidence for the formation of WC domains in the regime of integer[16,17] and fractional[10,11] $\nu$ is provided by the observation of resonances in the microwave conductivity. These resonances are interpreted as arising from shear modes of WC domains pinned by disorder and, consequently, as a manifestation of a finite shear modulus[3,10,14], a distinguishing feature of a solid. We use NMR to demonstrate another defining feature of a solid related to its structural properties. The Knight shift measures the effective magnetic field that electron spins exert on the nuclei of the host material[18], making it a sensitive probe of the local electron density and its spatial modulation.

We studied a 2DES confined to a 27-nm-wide GaAs quantum well using a resistively detected NMR (RD-NMR) technique[19] in the millikelvin regime (see Methods). Figure 1a shows resonance spectra of $^{75}$As nuclei at $B = 6.4$ T, obtained at various filling factors around $\nu = 2$, accessed using a gate voltage. At $\nu = 2$, equal numbers of spin-up and spin-down electrons fill the lowest $N = 0$ LL, resulting in zero net spin polarisation. The corresponding resonance spectrum thus represents the bare resonance frequency of the $^{75}$As nuclei, which is unaffected by the electron spin. As we move away from $\nu = 2$, the addition of spin-up electrons to the first excited $N = 1$ LL or removal of spin-down electrons from the filled $N = 0$ LL generates a partial filling factor $\nu^* = |\nu - 2|$ with a finite net spin polarisation (Fig. 1b), which shifts the spectra to lower frequencies (Knight shift).

In a simple picture, the net spin polarisation is proportional to the number of electrons or holes added to the $\nu = 2$ background, so that the Knight shift $K_s$ should increase linearly with $\nu^*$. Notably, we observe an atypical nonlinear suppression of $K_s$ at small $\nu^*$ accompanied by striking spectral anomalies. This becomes evident by comparing the experimentally observed spectra with simulations derived from a uniform 2DES. The simulations take into account the variation of the local electron (hole) density $\rho(z) = n^*|\psi_\nu(z)|^2$ along the direction normal to the plane of the 2DES, with $\psi_\nu(z)$ being the subband wave function (Fig. 1c) and $n^*$ ($\equiv \nu^* eB/h$) the sheet density of the electrons or holes added to $\nu = 2$. The simulation, which assumes the system to be uniform along the $x$-$y$ plane, well reproduces the spectra at $\nu = 2 - 1/3$ and $\nu \geq 2 + 1/5$ (solid lines). These "normal" lineshapes, characterised by sharp low-frequency onsets and elongated high-frequency tails, can be understood as a spectroscopic map of $|\psi_\nu(z)|^2$.[18, 19] However, the simulation fails to reproduce the spectra observed at $2 - 1/3 < \nu < 2$ and $2 < \nu < 2 + 1/5$ (dotted lines). $K_s$, measured at the peak of the spectra, is substantially below that expected from the simulation. Concurrently, we observe striking spectral anomalies, characterised by an elongated low-frequency tail and a high-frequency cutoff, which are most pronounced at $\nu = 1.8$ and $2.1$. Intriguingly, around $\nu = 2.1$, the low-frequency tail exceeds the $K_s$ value for a maximally polarised uniform 2DES.

We can explain both the atypical spectral lineshapes and the strong suppression of $K_s$ coherently if we assume that the electrons have solidified into a periodic lattice. Figure 2a depicts $\rho(x, y)$, the calculated spatial variation of the local electron density in the $x$-$y$ plane for a Wigner solid. The calculation for $\nu < 2$ assumes the Maki-Zotos ansatz (MZA) wave function[14], in which a hole in the filled $N = 0$ LL, described by a single-particle harmonic-oscillator wave function, is localised at each lattice site. For $\nu > 2$ we extended the MZA by replacing the single-particle wave function with that for the $N = 1$ LL. The MZA is a mean-field Hartree-Fock solution, which describes an uncorrelated WC at $T = 0$ K. We therefore introduced a tunable parameter $\sigma$, which controls the Gaussian blurring that is added to the in-plane density variation via the function $G_\sigma(x, y) \propto \exp[-(x^2 + y^2)/\sigma^2]$, to model correlation and finite-temperature effects. The spatially varying probability density of a WC induces a spatially varying Knight shift which deforms the NMR spectrum. By taking into account both the in-plane and out-of-plane density variation given by ρ(x, y) and $|\psi_v(z)|^2$, respectively, we are able to simulate the NMR spectrum for a Wigner solid (see Supplementary Information). Figure 2b demonstrates that the simulations with σ in the range from 0 to ~$\ell_B$ faithfully reproduce the anomalous NMR spectra observed around ν = 1.8 and 2.1.

The spectra provide far reaching insight into the internal architecture of the solid, such as the quantum properties of its constituent particles. In the lower panels of Fig. 2c, the measured $K_s$ is plotted against $\nu^*$ and compared with calculations using σ = 0 and $\ell_B$ (thin dashed and solid lines, respectively). In contrast to the conduction onset in the N = 0 and N = 1 LLs situated at similar $\nu^*$ values, as manifested by a rise in the longitudinal resistance (Fig. 2c, upper panels), the transition from a solid to a liquid phase occurs at much smaller $\nu^*$ (~ 0.15) in the N = 1 LL than in the N = 0 LL ($\nu^*$ ~ 0.28), reflecting the larger spatial extent of the N = 1 wave function. Furthermore, the low-frequency shoulder, peculiar to N = 1 LL spectra such as at ν = 2.1, is a hallmark of the nodal structure of the second LL wave function. These characteristics, determined by Landau orbital index, highlight the quantum constitution of the WCs and their liquid-solid phase competition. Our simulations show that the low-frequency shoulder at ν = 2.1 is sensitive to the σ value (see Supplementary Information), which in turn allows us to quantify the deviation from the T = 0 Hartree-Fock WC.

Figure 3a illustrates the spectral evolution at elevated temperatures. While the spectra shift to lower frequencies, with the peak approaching the position expected for a uniform electron liquid, the data clearly show that the WCs have not yet melted even at 350 mK. The evolution of the spectra can be reproduced by exploiting σ as a temperature-dependent parameter that describes the particles' displacement from their equilibrium positions, induced by thermally excited phonon modes. Physically, $\sigma^2$ represents the *extra* mean square displacement that is added to the zero-point fluctuation at T = 0, which is given by $\lambda^2 = 2\ell_B^2$ and $4\ell_B^2$ for the uncorrelated WC in the N = 0 and N = 1 LLs, respectively (see Supplementary Information). In Fig. 3b, we plot $(\sigma/a_v)^2$ versus temperature for $\nu$ = 1.8, 1.9 and 2.1, obtained from the fits, where $a_v = (2/\sqrt{3}n^*)^{1/2}$ is the lattice constant. These values can be translated into the Lindemann parameter $\langle u^2 \rangle/a_v^2 \equiv (\lambda^2 + \sigma^2)/a_v^2$, as indicated on the right axis. Consistent with theory[13], $(\sigma/a_v)^2$ increases almost linearly with $T$. However, the data reveal the temperature coefficients to be significantly greater than expected from the simple model[13], assuming a realistic size ($L \approx$ 0.5-2 μm)[10,20] of the WC domains. Such deviation can be attributed to the finite thickness of the 2DES, which weakens the Coulomb interaction and thus reduces the magneto-phonon frequency through the shear modulus[21]. For $\nu$ = 2.1, a linear extrapolation finds a finite $\sigma$ of ~ $0.8\ell_B$ at $T$ = 0 K, which suggests correlation effects that require a comparison with more elaborate theories[15,22-24].

The capability of NMR to probe the internal structure of a 2DES also allows us to identify the origin of the insulating phase at low $\nu$. At $\nu$ < 1/3, where transport measurements indicate that the system enters an insulating phase, NMR shows spectral anomalies (Fig. 4a) and strong suppression of $K_s$ (Fig. 4b), which are found to be very similar to those observed for 5/3 < $\nu$ < 2 if particle-hole symmetry $\nu \leftrightarrow 2 - \nu$ is invoked. Thus, we can say unambiguously that the insulating behavior at $\nu$ < 1/3 is due to electron solidification, and not disorder-induced single-particle localisation of an Anderson-type insulator[25]. Moreover, the behavior of $K_s$ versus $\nu$ clearly shows that, under our experimental conditions, the localised particles in the WC are electrons and not fractionally charged quasiparticles which would appear at higher fields and in cleaner samples[26]. The existence of a solid phase at $\nu$ = 1/5 and in the vicinity of $\nu$ = 1/3 is not trivial. Theories demonstrate that, in the thermodynamic limit, the Laughlin liquid state, being stabilized by long-range quantum fluctuations, is the ground state at these $\nu$ values[15,22-24]. In real systems, however, disorder may introduce finite-size effects into the 2DES, where a correlated crystalline phase can have a lower energy at $\nu$ = 1/5.[27] Figure 4b reveals that, as $\nu$ increases above 1/5, the system starts to deviate rapidly from the uncorrelated WC, which is indicated by the increase in $\sigma$ (Fig. 4b inset). This continues until the system gives way to the liquid phase at $\nu$ = 1/3. Our data thus constitute spectroscopic evidence for the evolution of quantum electron solids driven by the interplay between disorder and quantum fluctuations.

NMR spectroscopy of 2DESs has mostly been used to probe the electron spin polarisation via the Knight shift. Our NMR experiments have demonstrated that it is a far more powerful tool, capable of revealing nanoscale details of the spatial variation of spin/charge density that emerges when the translational symmetry is broken. This work provides the framework for future experiments that probe other exotic phases with non-trivial spatial spin/charge order, including stripe/bubble phases[28], re-entrant insulating phase[29], and quantum Hall nematic phases[30].

## Methods

The sample is a 100-µm-wide Hall bar with the 2DES confined to a 27-nm-wide gallium arsenide (GaAs) quantum well, grown on a $n^+$-substrate that serves as back gate. The sample is inside the mixing chamber of a dilution refrigerator and surrounded by a three-turn coil which is connected to a frequency generator. Resonance spectra were obtained, following Ref. 19, by monitoring the changes in the longitudinal resistance $R_{xx}$ near $\nu_{read}$ = 0.59, induced by radio frequency excitations (–17 dBm) at the filling factors of interest (i.e., $\nu = 2 \pm \nu^*$ and $\nu < 1/3$).

# References

1. Tsui, D. C., Stormer, H. L. & Gossard, A. C. Two-dimensional magnetotransport in the extreme quantum limit. *Phys. Rev. Lett.* **48,** 1559–1562 (1982).
2. Wigner, E. P. On the interaction of electrons in metals. *Phys. Rev.* **46,** 1002–1011 (1934).
3. Andrei, E. Y. *et al.* Observation of a magnetically induced Wigner solid. *Phys. Rev. Lett.* **60,** 2765–2768 (1988).
4. Willett, R. L. *et al.* Termination of the series of fractional quantum hall states at small filling factors. *Phys. Rev. B* **38,** 7881 (1988).
5. Jiang, H. W. *et al.* Quantum liquid versus electron solid around ν = 1/5 Landau-level filling. *Phys. Rev. Lett.* **65,** 633–636 (1990).
6. Goldman, V. J., Santos, M., Shayegan, M. & Cunningham, J. E. Evidence for two-dimensional quantum Wigner crystal. *Phys. Rev. Lett.* **65,** 2189–2192 (1990).
7. Williams, F. I. B. *et al.* Conduction threshold and pinning frequency of magnetically induced Wigner solid. *Phys. Rev. Lett.* **66,** 3285–3288 (1991).
8. Buhmann, H. *et al.* Novel magneto-optical behavior in the Wigner-solid regime. *Phys. Rev. Lett.* **66,** 926–929 (1991).
9. Kukushkin, I. V. *et al.* Evidence of the triangular lattice of crystallized electrons from time resolved luminescence. *Phys. Rev. Lett.* **72,** 3594–3597 (1994).
10. Ye, P. D. *et al.* Correlation lengths of the Wigner-crystal order in a two-dimensional electron system at high magnetic fields. *Phys. Rev. Lett.* **89,** 176802 (2002).
11. Chen, Y. P. *et al.* Melting of a 2D quantum electron solid in high magnetic field. *Nat. Phys.* **2,** 452–455 (2006).
12. Lozovik, Yu. E. & Yudson, V.I. Crystallization of a two-dimensional electron gas in a magnetic field. *JETP Lett.* **22,** 11 (1975).
13. Ulinich, F. P. & Usov, N. A. Phase diagram of a two-dimensional Wigner crystal in a magnetic field. *Zh. Eksp. Taor. Fiz.* **76,** 288-294 (1979).
14. Maki, K. & Zotos, X. Static and dynamic properties of a two-dimensional Wigner crystal in a strong magnetic field. *Phys. Rev. B* **28,** 4349–4356 (1983).

15. Lam, P. K. & Girvin, S. M. Liquid-solid transition and the fractional quantum-Hall effect. *Phys. Rev. B* **30,** 473 (1984).
16. Chen, Y. *et al.* Microwave resonance of the 2D Wigner crystal around integer Landau fillings. *Phys. Rev. Lett.* **91,** 016801 (2003).
17. Zhu, H. *et al.* Pinning-mode resonance of a Skyrme crystal near Landau-level filling factor $\nu = 1$. *Phys. Rev. Lett.* **104,** 226801 (2010).
18. Kuzma, N. N., Khandelwal, P., Barrett, S. E., Pfeiffer, L. N., West, K. W. Ultraslow electron spin dynamics in GaAs quantum wells probed by optically pumped NMR. *Science* **281,** 686–690 (1998).
19. Tiemann, L., Gamez, G., Kumada, N., Muraki, K. Unraveling the spin polarization of the $\nu = 5/2$ fractional quantum Hall state. *Science* **335,** 828–831 (2012).
20. Martin, J. *et al.* Localization of fractionally charged quasi-particles. *Science* **305,** 980–983 (2004).
21. Ettouhami, A. M., Klironomos, F. D. & Dorsey, A. T. Static and dynamic properties of crystalline phases of two-dimensional electrons in a strong magnetic field. *Phys. Rev. B* **73,** 165324 (2006).
22. Yi, H. & Fertig, H. A. Laughlin-Jastrow-correlated Wigner crystal in a strong magnetic field. *Phys. Rev. B* **58,** 4019–4027 (1998).
23. Shibata, N. & Yoshioka, D. Ground state phase diagram of 2D electrons in high magnetic field. *J. Phys. Soc. Jpn.* **72,** 664–672 (2003).
24. Chang, C.-C., Jeon, G. S., & Jain, J. K. Microscopic verification of topological electron-vortex binding in the lowest Landau-level crystal state, *Phys. Rev. Lett.* **94,** 016809 (2005).
25. Kivelson, S., Lee D.-H., Zhang, S.-C. Global phase diagram in the quantum Hall effect. *Phys. Rev. B* **46,** 2223–2238 (1992).
26. Zhu, H. *et al.* Observation of a pinning mode in a Wigner solid with $\nu = 1/3$ fractional quantum Hall excitations. *Phys. Rev. Lett*. **105,** 126803 (2010).
27. Chang, C.-C., Töke, C., Jeon, G. S., & Jain, J. K. Competition between composite-fermion-crystal and liquid orders at $\nu = 1/5$. *Phys. Rev. B* **73,** 155323 (2006).

28. Lilly, M. P., Cooper, K. B., Eisenstein, J. P., Pfeiffer, L. N., & West, K. W. Evidence for an anisotropic state of two-dimensional electrons in high Landau levels. *Phys. Rev. Lett.* **82,** 394–397 (1999).
29. Eisenstein, J. P., Cooper, K. B., Pfeiffer, L. N., & West, K. W. Insulating and fractional quantum Hall states in the first excited Landau level. *Phys. Rev. Lett.* **88,** 076801 (2002).
30. Xia, J., Eisenstein, J. P., Pfeiffer, L. N. & West, K. W. Evidence for a fractionally quantized Hall state with anisotropic longitudinal transport. *Nat. Phys.* **7,** 845 (2011).

## **Acknowledgements**

The authors would like to acknowledge Rudolf Morf for insightful discussions and valuable comments. Appreciation also goes to Norio Kumada for experimental support and Tatsuya Higashi for sharing the results of his theoretical calculations.

# Figure legends

**Figure 1. Evolution of RD-NMR spectra with filling factor. a**, RD-NMR spectra of $^{75}$As nuclei measured at 6.4 T and base temperature in the range $5/3 < \nu < 7/3$. Spectra taken at different $\nu$ are vertically offset for clarity. The left (right) panel shows spectra for $\nu \leq 2$ ($\nu \geq 2$). The $\nu = 2$ spectrum is fitted with a Gaussian function, which provides the bare resonance frequency. Experimental data (markers) are compared with numerical simulations using a model based on a uniform 2DES (filled area lines). At $\nu = 5/3$ and $\nu \geq 2.2$, experimental spectra closely match simulations, which are delimited by solid lines. Experimental data match poorly with simulations at other $\nu$ (delimited by dashed lines). The electron temperature under radio-frequency irradiation is estimated to be 50 mK. **b**, Schematic illustrations of occupied Landau levels for $\nu < 2$ and $\nu > 2$. **c**, Confinement potential (gray) and squared subband wave function (pink). The latter represents the out-of-plane variation of electron density in the quantum well.

**Figure 2. Comparison of experiment with a model incorporating a spatially varying density landscape of a Wigner solid. a**, Top panel: In-plane variation of local density $\rho(x, y)$ calculated for Wigner solids at $\nu = 1.9$ and 2.1 for zero $\sigma$ (uncorrelated WC) and finite $\sigma$ (blurred WC). Bottom panel: corresponding local density profile along the lattice sites in the $x$ direction. The density profile for $\nu = 1.8$ is also displayed. The vertical axes are scaled with respect to the density of a uniform electron system at each $\nu$ shown by the horizontal dotted lines. **b**, Fitting of RD-NMR spectra at 6.4 T for several $\nu$ in the $N = 0$ and $N = 1$ LLs. Experimental data (markers) are compared with numerical fits (filled-area lines) using a theoretical model incorporating the presence of WC domains. The fitting parameter $\sigma$ is displayed for each $\nu$. **c**, Upper panels: Longitudinal resistance $R_{xx}$ as a function of effective filling factor $\nu^*$ for the $N = 0$ and $N = 1$ LLs. Lower panels: Knight shift $K_s$ measured at the peak of the spectra as a function of $\nu^*$ for varying LL index. The vertical bars represent the half width at half maximum of the spectra on each side of the peak. Experimental data (markers) are compared with the expected $\nu$ dependence for a

uniform 2DES (solid blue lines), that of a WC with $\sigma = 0$ (dashed black lines) and that of a WC with $\sigma = \ell_B$ (solid black lines).

**Figure 3. Temperature dependence of RD-NMR spectra. a**, Temperature evolution of RD-NMR spectra at $\nu = 1.8$, 1.9 and 2.1 at 6.4T. Experimental data (markers) are compared with numerical fits based on a model incorporating the WC and finite $\sigma$ (filled-area lines) and simulations based on a uniform 2DES (dashed black lines). Spectra taken at different temperatures are vertically offset for clarity. For ease of comparison, spectra are normalised with respect to the simulation for a uniform 2DES. **b**, Temperature dependence of the parameter $\sigma$ obtained for $\nu = 1.8$, 1.9 and 2.1 from the fits shown in **a**. The left axis is $\sigma^2$ in units of $a_\nu^2$, with $a_\nu$ the lattice constant at each $\nu$, while the right axis is the Lindemann parameter. Markers represent the values obtained from the fits shown in **a**, with the error bars indicating the variation induced by a shift in the reference frequency by $\pm 0.3$ kHz. Thick gray lines indicate linear fits to the $\sigma^2$ vs $T$ plot, obtained with the constraint $\sigma(T = 0) \geq 0$. Calculations based on the model in Ref. 13 (filled orange area), assuming a WC domain size of 0.5-2 μm, are displayed for comparison.

**Figure 4. Evidence of Wigner solids from NMR spectra at $\nu < 1/3$. a**, RD-NMR spectra (markers) at 6.4 T and base temperature for several $\nu \leq 1/3$. The spectrum at $\nu = 1/3$ closely matches the simulation assuming a uniform 2DES (blue line). The spectra at $\nu < 1/3$ can be fitted well with a model incorporating the WC and finite $\sigma$ (pink lines). **b**, Main panel: Knight shift $K_s$ (markers) measured at the peak of the spectra, plotted as a function of $\nu$. The vertical bars represent the half width at half maximum of the spectra on each side of the peak. Experimental data are compared with the expected behavior of a uniform 2DES (solid blue line), that of an uncorrelated WC (dashed pink line) and that of a WC with $\sigma = \ell_B$ (solid pink line). The black dot-dashed line is the predicted dependence assuming that the WC comprises not electrons but instead Laughlin quasiholes. Inset: Parameter $\sigma$ obtained from the fits shown in **a**, normalized by $\ell_B$ and plotted as a function of $\nu$. Error bars indicate the variation induced by a shift in the reference frequency by $\pm 0.3$ kHz.

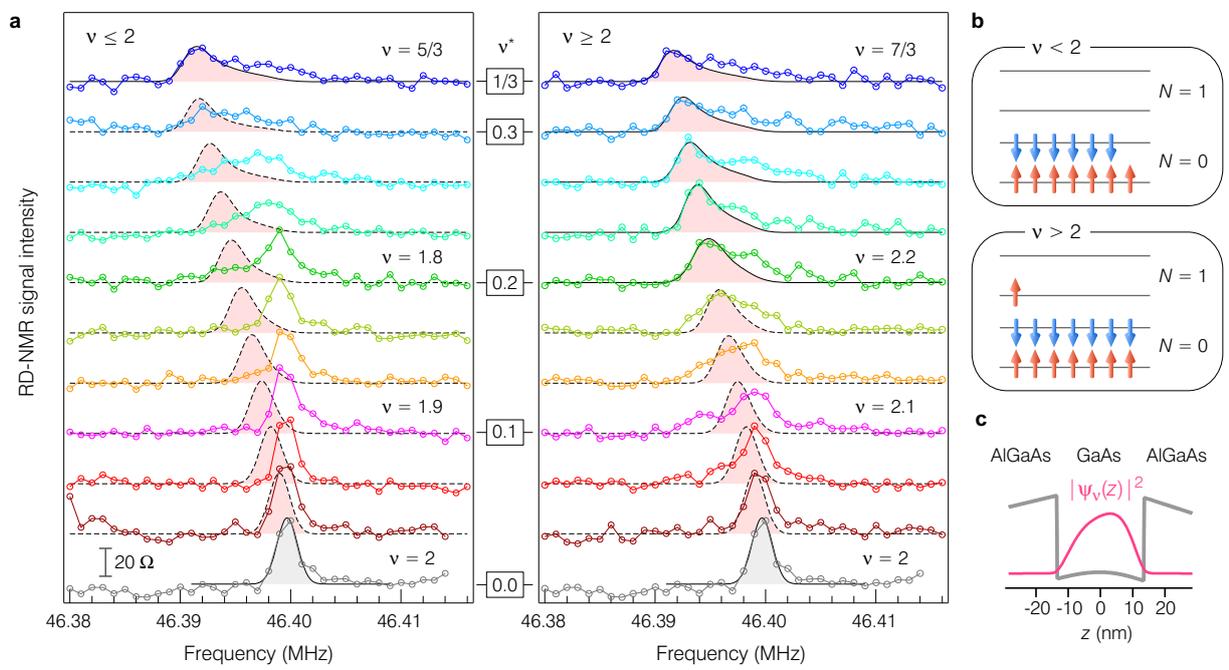

Fig. 1

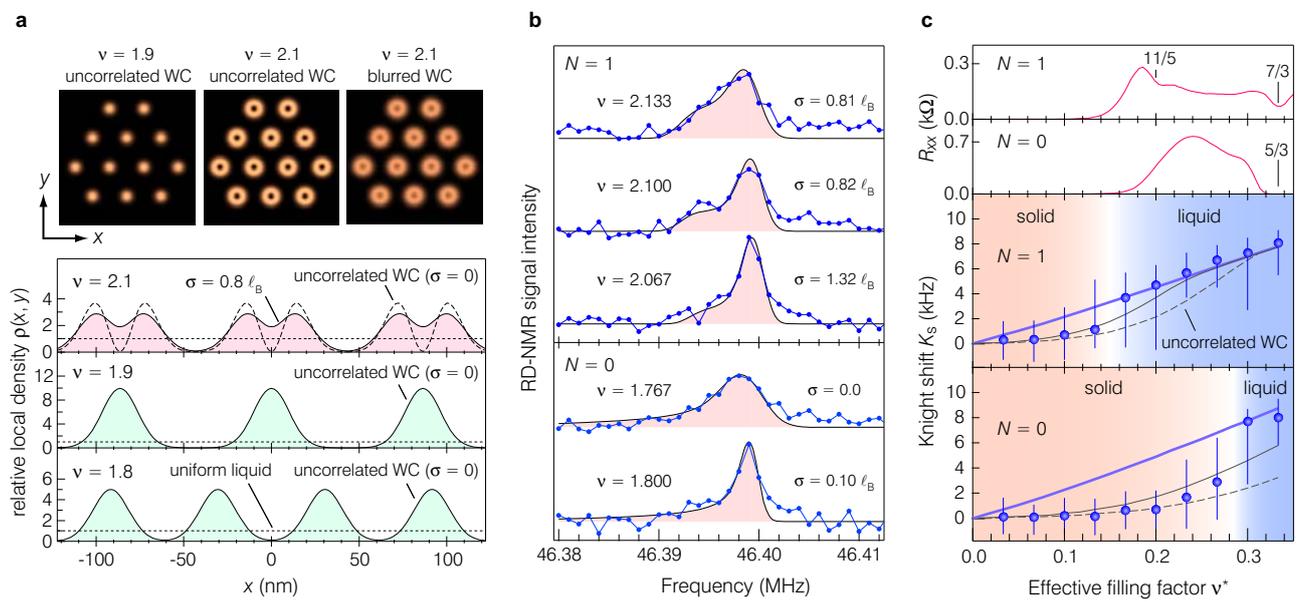

Fig. 2

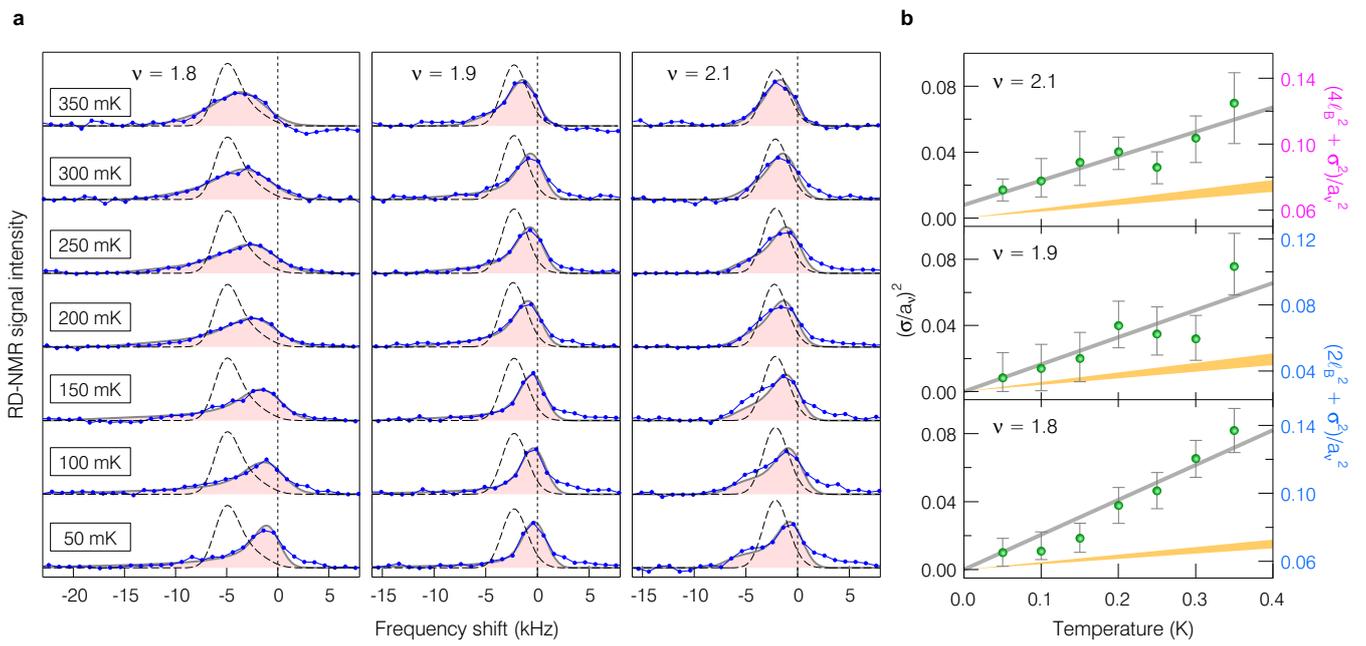

Fig. 3

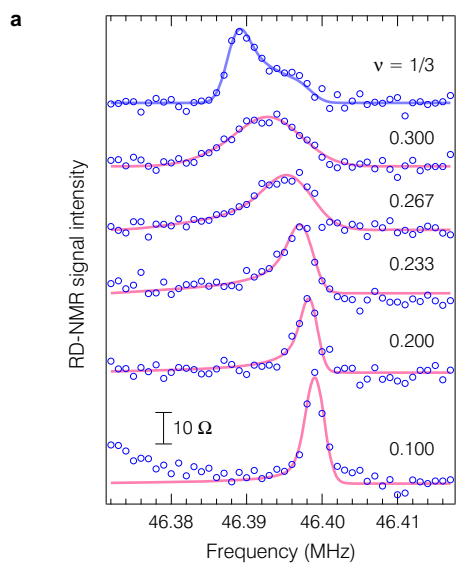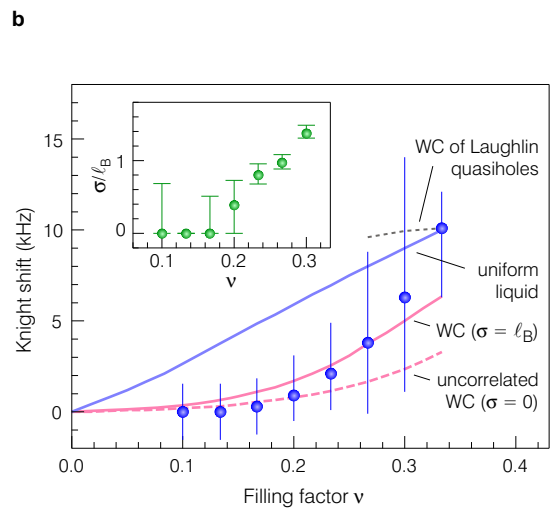

Fig. 4